\documentclass[twocolumn,showpacs,preprintnumbers,amsmath,amssymb]{revtex4}

\usepackage{graphicx}
\usepackage{dcolumn}
\usepackage{bm}

\begin{document}


\title{Quantum Gravity as a Network\\Self-Organization of a Discrete 4D Universe}


\author{Carlo A. Trugenberger}
\email{ca.trugenberger@bluewin.ch}
\affiliation{%
SwissScientific, chemin Diodati 10, CH-1223 Cologny, Switzerland
}%

\date{\today}

\begin{abstract}
I propose a quantum gravity model in which the fundamental degrees of freedom are information bits for both discrete space-time points and links connecting them. The Hamiltonian is a very simple network model consisting of a ferromagnetic Ising model for space-time vertices and an antiferromagnetic Ising model for the links. As a result of the frustration between these two terms, the ground state self-organizes as a new type of low-clustering graph with finite Hausdorff dimension 4. The spectral dimension is lower than the Hausdorff dimension: it coincides with the Hausdorff dimension 4 at a first quantum phase transition corresponding to an IR fixed point while at a second quantum phase transition describing small scales space-time dissolves into disordered information bits. The large-scale dimension 4 of the universe is related to the upper critical dimension 4 of the Ising model.  At finite temperatures the universe graph emerges without big bang and without singularities from a ferromagnetic phase transition in which space-time itself forms out of a hot soup of information bits. When the temperature is lowered the universe graph unfolds and expands by lowering its connectivity, a mechanism I have called topological expansion. The model admits topological black hole excitations corresponding to graphs containing holes with no space-time inside and with ``Schwarzschild-like" horizons with a lower spectral dimension. 
\end{abstract}
\maketitle

\section{Introduction}
Three main approaches to the problem of quantum gravity have crystallized over the last thirty years. Superstring theory \cite{strings} was introduced to bundle the graviton and gauge vector bosons in a single consistent mathematical framework. This, unfortunately requires many additional, as yet unobserved degrees of freedom as well as new symmetries and dimensions. It appears by now, however that strings themselves have to embedded in a larger framework, going under the name of M-theory \cite{mother}. The major problem of M-theory, however is that it is not even properly defined at the non-perturbative level.

Loop quantum gravity \cite{loop}, instead, is a canonical approach to the Einstein-Hilbert action formulated in terms of connections and vielbeins, analogously to the canonical quantization of gauge theories. Correspondingly, the main variables are holonomies (loops).  Despite its many successes and appealing structure at the Planck length \cite{baez}, however, loop quantum gravity has not yet been able
to successfully predict the geometries expected in the semi-classical limit. 

Finally, the Wilsonian approach based on causal dynamical triangulations (CDT) \cite{triang} is the gravity equivalent of lattice gauge theories. It is formulated by discretizing space-time in terms of (generalized) triangles, formulating the gravitational action via a modified Regge calculus \cite{regge} and summing over all possible triangulations to obtain the quantum theory. The idea is to look for non-Gaussian ultraviolet fixed (UV) points that would define quantum gravity in a non-perturbative sense, an idea going under the name of asymptotic safety \cite{safety}. There is indeed compelling numerical evidence for the emergence of a de Sitter universe with finite Hausdorff dimension and scale-dependent spectral dimension \cite{spec}. The physics of these solutions is suggestive of a deep relation with a recent proposal \cite{hor} of a quantum gravity model in which time and space are fundamentally different at high energies, with the Lorentz symmetry emerging only as a low-energy effective symmetry. This ties in with the failure of an earlier approach based on Euclidean dynamical triangulations, with no imposed causal structure \cite{edta}. Euclidean dynamical triangulations have been recently revisited \cite{edtb} but no improvements upon CDTs have been found. 

Despite their differences, all these approaches share the common principle of starting from the correct classical Einstein-Hilbert gravity action and trying to reconcile it with quantum mechanics, either by enlarging the Hilbert space, by using different quantum variables or by discretizing the degrees of freedom. In this paper I will advocate a different point of view by ascribing the non-renormalizability of perturbative quantum gravity to a complete misidentification of the fundamental degrees of freedom. I will suggest that geometric variables are not suited to formulate quantum gravity since the fundamental degrees of freedom are purely combinatoric and describe nothing else than information bits. In this picture, the geometry of our universe is an emergent property of the quantum ground state at a critical infrared (IR) fixed point while the small-scale physics is governed by an ultraviolet fixed point for information bits with no geometric space-time interpretation. Note that this is essentially the asymptotic safety scenario in disguise: in the usual approach one starts from the correct infrared fixed point corresponding to general relativity and one looks for a non-trivial ultraviolet fixed point. Here I propose the reverse approach, to start from a different set of microscopic degrees of freedom describing the UV physics and 
trying to recover the existence of a flow to an infrared fixed point corresponding to general relativity. 

Of course I will not be able to complete this program in the present paper. However, I will try
to substantiate my claim by introducing a surprisingly simple model of a network in which a 4-dimensional (4D) discrete universe is the ground state that emerges spontaneously in one of the three quantum phases of the model. This emerging discrete universe is embodied by a low-clustering \cite{graphrev} graph with finite Hausdorff dimension 4. As in the CDT approach, the spectral dimension of this universe graph is smaller than its Hausdorff dimension. Admittedly, in this paper I concentrate only on the first step, from combinatorics to topology, geometry and the link to general relativity and the Einstein-Hilbert action being still missing. Nonetheless, I believe the phase structure of the emerging topology is very suggestive of a connection with quantized space-time. 

The model I propose is a generalization of Kazakov's random lattice planar (2D) Ising model \cite{kazakov} to include dynamical topologies, governed by a second, antiferromagnetic Ising model for edges of a random graph. In a nutshell, this second, edge Ising model renders the connectivity of the original Ising model dynamical by making it dependent on the dimensionless coupling constant. In Kazakov's model the local limit at the critical Ising temperature describes 2D quantum gravity  coupled to Majorana fermions, as can be derived exactly by matrix model techniques \cite{alvarez}. In the present model instead, the two relevant phase transitions do not depend on a critical temperature but are purely quantum. Here, the ground state of the original Ising model reacts to changes of the dynamical connectivity $2d$, which coincides with the dimensionless coupling constant of the model. Below the lower critical dimension $d_{\rm lc}=1$ the Ising model is disordered: correspondingly there is a first quantum phase transition at $d=1$ in which the network breaks up into many small, disconnected and disordered components and any space-time interpretation is lost, the "universe" dissolving in fluctuating information bits. At the upper critical dimension $d_{\rm uc}=4$ the character of the Ising model ground state changes again. In this phase the Hausdorff dimension of the "universe graph" coincides with its spectral dimension. Between these two quantum transitions there is a phase in which the spectral dimension is lower than the Hausdorff dimension, $d_s < d_H$ and the numerical evidence strongly points to a Hausdorff dimension taking the unique fixed value $d_H=4$, this being also supported by recent analytic results on the incommensurability between the physical and the correlation length in finite size systems above the upper critical dimension \cite{berche}. 

I cannot compute analytically the order of the two quantum phase transitions. However I conjecture that they correspond to UV ($d$=1) and IR ($d$=4) critical points and realize thus the asymptotic safety scenario. The large scale dimension 4 of the universe would thus be predicted and related to the upper critical dimension of the Ising model. The UV critical point at $d$=1 would, additionally, imply that space-time ceases to exist at very small scales.  As in the CDT approach \cite{triang} the spectral dimension of the universe decreases at small scales, but it does not stop at d=2, rather it goes all the way down to the Ising lower critical dimension $d=1$ where space-time ceases to exist. 

The model can be treated at finite temperatures ($T$) in mean field theory. Of course, here one cannot think of temperature in terms of time averages of microstates since there is no time yet. Temperature, rather, has to be considered as a further external parameter/coupling constant that assigns non-vanishing probabilities to graphs different than the ground state according to their energies, something like the stochastic temperature \cite{huf} associated with the Parisi-Wu fictitious time of stochastic quantization \cite{pawu}. As I will show, this parameter can induce anisotropies in the universe graph, a most interesting possibility to distinguish a preferred direction. 

Mean field theory, however is unsuited to decide if the two quantum phase transitions are isolated or, rather, the endpoint of continuous lines of phase transitions at finite $T$. On the contrary, it should be quite accurate for high connectivities. Here it predicts a ``topological cosmic expansion" due to a decrease in the graph connectivity when the temperature is lowered, i.e. the universe graph actually ``unfolds" in this phase rather than expand. There is no ``big bang singularity", the origin of the universe graph is a ferromagnetic phase transition in which the space-time information bits align and connect, i.e. discrete space-time itself self-organizes. Above the critical temperature there is only a hot information soup with maximal entropy and no possible interpretation as discrete space-time. During topological expansion the average distance on the graph scales exponentially with the temperature. A small drop in temperature causes a completely connected universe graph to unfold to a regular graph of dimension 4. 

Finally, macroscopic ``topological black holes" appear naturally as excitations of the model. It can be shown that such configurations are stable down to very low temperatures. 

The idea that in a proper theory of quantum gravity space-time should emerge spontaneously is not new. Models in which space is dynamically generated in quantum models with a well-defined notion of time were already discussed in the 80s \cite{wein}. The same idea is also at the basis of the causal set approach to quantum gravity \cite{sorkin} and of its recent development into energetic causal sets \cite{smolin}. Ideas similar to those introduced in this paper have been previously considered in the so-called quantum graphity model \cite{graphity}. The Hamiltonian of this model, however, is substantially different (and more complex) from the one proposed here and the derivation of the self-organization of a 4D universe-like graph from information combinatorics is entirely missing. Another related approach is based on structurally dynamic cellular networks (SDCNs) and is reviewed in in \cite{sdcn}. Finally, let me mention a recently developed, related approach \cite{bianconi} based on the idea of constructing discrete space-times by growing networks of simplicial complexes.

\section{The model}
Consider $N$ spin 1/2 information bits $s_i = \pm 1$, for $i=1\dots N$ and $N(N-1)/2$ spin 1/2 information bits $w_{ij}=0,1$, for
$i,j=1\dots N$. A value of $s_i=+1$ denotes the existence of space time, while $s_i=-1$ indicates the absence of space time (or the presence of anti-space-time). A value $w_{ij}=1$ denotes a connection between bits $s_i$ and $s_j$, a value $w_{ij}=0$ indicates that the two spins $s_i$ and $s_j$ are not connected. 

All these information bits form a network with energy
\begin{equation}
H = {J\over 2} \ \sum_{i\ne j} \sum_{k\ne i \atop k\ne j} w_{ik}w_{kj} - {1\over 2} \sum_{i\ne j} s_i w_{ij} s_j \ , 
\label{twob}
\end{equation}
where the link variables are symmetric, $w_{ij} =w_{ji}$ and vanish on the diagonal, $w_{ii} = 0$ and where $J$ is the unique dimensionless coupling. I use units in which $\hbar=1$, $c=1$ and all energies are measured in units of the standard Ising coupling, second term in (\ref{twob}), which is set to one for simplicity of presentation. 

The second term in this energy function is the standard ferromagnetic Ising model. If the links $w_{ij}$ would be uniformly drawn from random adjacency matrices of degree 4, the model would be Kazakov's random lattice Ising model in two dimensions \cite{kazakov}. The first term in the energy functional, on the other side, is simply a nearest-neighbours (sharing a common vertex) antiferromagnetic Ising model for the link spins. The generalizations with respect to Kazakov's model, thus consist in dropping the restriction to degree 4 and drawing the random adjacency matrices from a Gaussian distribution. 

The competition between the vertex ferromagnetic coupling and the link antiferromagentic one creates "link frustration" in the model. Indeed the vertex ferromagnetic coupling favours the creation of many links (positive values of $w_{ij}$) in a state with the majority of vertex spins aligned, corresponding to an incipient space-time: space-time points have a tendency to link together. However, due to the antiferromagnetic link coupling, creating many links costs energy and is energetically disfavoured. It is particularly nearest neighbour links, sharing a common vertex, that are disfavoured. As I now show, the compromise is to create links between space-time points but to avoid triangles involving nearest neighbours links.  The result is a graph with power-law extension, exactly what one would expect for a discretized universe. The dimension of this universe, encoded in the power-law for the average distance is determined by the unique dimensionless coupling $J$ of the model. 

To show this, let me consider a $k$-regular graph, i.e. a configuration with $s_i = +1$, $\forall i$ and such that each vertex has exactly $k$ incident edges (degree $k$ in graph parlance \cite{graphrev}). When $k$ is even, such a graph resembles locally a $k/2$-dimensional lattice. I will now show that such a graph is a local minimum of the energy (\ref{twob}). 

To this end, suppose we are given such a graph and let me first change one single vertex spin from $s_i=+1$ to $s_i=-1$. The corresponding energy change $\Delta E_i = k > 0$ is positive. Changing vertex spins, thus costs energy. Let us now try to eliminate an existing connection, i.e. changing $w_{ij} = 1$ to $w_{ij} = 0$. In this case there are contributions from both terms in the energy function. Before the elimination, the existing connection contributed $E_{ij} = J(k-1)-(1/2)$. After the elimination, of course the contribution of this connection vanishes. The energy change due to the elimination of a connection $w_{ij}$ is thus 
$\Delta_{\rm elim} E_{ij} = (1/2)- J(k-1)$. Let me now consider adding a previously non-existent connection $w_{ij}$. In this case it is the energy contribution before the addition that vanishes, while, after the addition I have an additional energy $E_{ij}= Jk -(1/2)$. The energy change for adding a connection is thus $\Delta_{\rm add} E_{ij} = Jk-(1/2)$. Requiring that both eliminating and adding a connection costs energy we obtain the stability condition
\begin{equation}
{1\over 2J} < k < 1+{1\over 2J} \ .
\label{twoc}
\end{equation}
To proceed, let me compute the total energy of a $k$-regular graph on $N$ vertices. This is easily obtained as
\begin{equation}
E_{N, k} = N \left( {J\over 2} k (k-1) -{k\over 2} \right) \ .
\label{twod}
\end{equation}
This expression is minimized when the vertex degree takes the value $k=1/2 + (1/2J)$ and this value of $k$ satisfies the stability condition (\ref{twoc}). Defining 
\begin{equation}
J = {1\over 4d-1} \ ,
\label{twoe}
\end{equation}
I have obtained the result that for any choice of integer $d$ a $2d$-regular graph is a local minimum of the energy and that this minimum is the one of lowest energy among all possible regular graphs. As anticipated, the links arrange themselves to form a locally lattice-like graph with $2d$ edges at each vertex; the number of triangles is minimized. Thus, $d$ plays the role of the "bare" dimension of the universe. I call this dimension "bare" since, as I will show below, the physical Hausdorff dimension of the universe can differ from $d$. 

Of course, I have not shown that $2d$-regular graphs are the true global minima of the energy functional (\ref{twob}). There could be other, non-regular graphs, with even lower energies. That this is not so can be made plausible with the numerical methods introduced in the next section. 

\section{The properties of emergent universe graphs} 

The energy function (\ref{twob}) defines a Maxwell-Boltzmann distribution on the space of graphs. This distribution is peaked around the ground state of $H$, which corresponds to the semi-classical ``universe graph". Graphs with large weights in the distribution correspond to fluctuations about this semi-classical universe graph. In the rest of this paper I shall focus on the properties of the semi-classical universe graph as a function of the dimensionless coupling $J$. 

To do so I will reverse a technique used to analyze neural networks such as the Hopfield model \cite{neural}. There, a network dynamics is posited and an energy functional is looked for that acts as a Lyapunov function, i.e. such that the energy minimum determines the fixed points of the network dynamics. Here I will do the contrary, namely I will look for a network dynamics whose fixed points correspond to minima of the energy (\ref{twob}). 

To this end let me start form a random initial configuration and sequentially update the vertex and link spins according to the rule
\begin{eqnarray}
&&s_i (t+1) = {\rm sign}_+ \left( h_i(t) \right) \ , \nonumber \\
&&h_i = \sum_{j\ne i} w_{ij} s_j  \ ,
\label{twof} \\
&&w_{ij}(t+1) = \Theta _{\pm} \left( h_{ij} \right) \, \nonumber \\
&&h_{ij} =  {1\over 2} s_i  s_j  - {J\over 2} \sum_{k\ne j \atop k\ne i} w_{kj}(t) - {J\over 2} \sum_{k\ne i \atop k\ne j} w_{ik}(t)  \ ,
\label{twog}
\end{eqnarray}
where $\Theta$ denotes the Heaviside function and $h_{ii}=0$. The subscripts "+" and "$\pm$" on the sign and $\Theta$ functions indicate what is the rule to follow in the (rare) cases in which the argument is zero. Complete symmetry would require a totally random choice in these cases. This, however can lead to a two-component network in which space-time vertices have the opposite sign in the two components. Since I am interested in the emergent properties of just one of these components I will maximize its size by making an asymmetric choice for the sign function determining the orientation of the space-time vertices. This is embodied in the subscript "+", which indicates that, in this case, the + sign has to be chosen. This favours positive vertex spin values so that one-component graphs are preferred. The subscript "$\pm$" on the Heaviside function, instead indicates that a purely random choice will be made when the argument vanishes. This choice also helps to avoid being stuck in sub-optimal minima during the evolution. 

I will now show that the energy function ({\ref{twob}) cannot increase along a sequential evolution (\ref{twof}) and (\ref{twog}): $E(t+1) \le E(t)$. To this end let me begin by considering the update of vertex spin $s_i$. The corresponding contribution $E_i$ to the energy changes according to
\begin{eqnarray}
E_i(t+1) &&= - s_i(t+1) \sum_{j\ne i} w_{ij} (t) s_j(t) 
\nonumber \\
&&= - {\rm sign}_+ \left( h_i(t) \right) h_i(t) =- |h_i(t)| 
\nonumber \\
&&\le -s_i(t) h_i(t) = E_i(t) \ .
\label{twoh}
\end{eqnarray}
With the exact same procedure for the update of a link spin $h_{ij}$ we obtain
\begin{equation}
E_{ij}(t+1) = - w_{ij} (t+1) h_{ij} (t) = -\Theta_{\pm} \left( h_{ij} (t) \right) h_{ij}(t) \ .
\label{twoi}
\end{equation}
Now we have two possibilities 
\begin{itemize} 
\item $h_{ij} (t) <0$ $\rightarrow$ $E_{ij}(t+1) =0 \le -w_{ij}(t)  h_{ij}(t) = E_{ij} (t)$ 
\item $h_{ij} (t) > 0$ $\rightarrow$ $E_{ij}(t+1) =- h_{ij}(t) = -|h_{ij} (t) | \le -w_{ij}(t)  h_{ij}(t) = E_{ij} (t)$ 
\end{itemize}
and in both cases we obtain $E_{ij} (t+1) \le E_{ij}(t)$, which proves the claim that the energy cannot increase during sequential evolution (note that, for both the sign and the Heaviside functions, the exact procedure on how to treat the undefined cases in which the argument vanishes has no effect on this result). As a consequence, every minimum of the energy (\ref{twob}) is a fixed point of the sequential network evolution (\ref{twof}) and (\ref{twog}). 

Every such fixed point defines an (undirected) graph. Several classes of graphs have been studied \cite{graphrev} extensively. The simplest such one is the class of regular graphs, for which every node has a fixed degree $k$, i.e. a fixed number $k$ of neighbours. Regular graphs are typically highly clustered, with the notable exception of $d$-dimensional hypercubic lattices, which have vanishing clustering coefficient since they do not contain any triangles, and in which the average distance scales as $N^{1/d}$. This is the exact opposite of classical random graphs \cite{randomgraphs} in which connections between nodes are chosen randomly with a given fixed probability. Random graphs have small clustering $c_{\rm rg}=\langle k \rangle/N$ where $\langle k \rangle$ is the degree expectation value and small average distance that scales logarithmically with $N$ (the so-called Òsmall worldÓ effect). In classical random graphs the degrees follow a Poisson distribution. In order to obtain graphs that better model real-world networks a class of generalized random graphs was introduced \cite{genrandomgraphs}, in which connections are chosen randomly subject to the constraint that the degree distribution must follow a predetermined law. In generalized random graphs the clustering depends, of course, from the chosen input degree distribution. Actually, only the first two moments of this distribution matter,
\begin{equation}
c_{\rm grg} = {1\over N} \ {\left( \langle k^2 \rangle -\langle k\rangle^2 \right)^2 \over \langle k\rangle^3} \ .
\label{grapha}
\end{equation}
This reduces to $c_{\rm rg}$ for a Poisson distribution, as expected. Generalized random graphs are also small worlds, the average distance remains a logarithmic function of $N$. These two classes describe static graphs, with a fixed number $N$ of nodes. The number of nodes grows, instead in the class of evolving graphs \cite{evrandomgraphs}. Many evolution rules have been considered \cite{graphrev}, essentially all aimed at generating scale-free graphs with power-law degree distributions of various forms. 

The graphs that minimize the energy (\ref{twob}) constitute yet another, novel category for which
\begin{itemize}
\item{} The number of nodes is not growing but fixed.
\item{} There is no random assignment of connections but, rather, these are dynamically determined.
The dynamics, however involves both connections {\it and} nodes. 
\item{} The dynamics is not an ad-hoc rule of how a graph is grown but is governed by an energy minimization principle. 
\item{} Contrary to generalized random graphs and typical evolving graphs, the degree distribution is not an input but an output. 
\end{itemize}
I shall call such graphs dynamical graphs and, in the following, I shall describe their main properties. 

In order to investigate these dynamical graphs, I have repeatedly started with random values of all the spins and let them evolve according to the dynamics (\ref{twof}) and (\ref{twog}) until a fixed point is reached. The properties of interest can then be extracted from the adjacency matrix of this graph. Finally, an average over the results form different runs is taken. Of course, care has to be taken to avoid ending up in fixed points corresponding to metastable, local minima of the energy function. Fortunately, these typically involve configurations for which not all space-time information bits take the same value $+1$ and can thus be recognized easily.  Actually, it could also be possible that a fixed point cannot be reached since the ground state is degenerate and so the network dynamics would wander among the different ground states with the same energy. This would be no problem however, since anyone of these ground states would correspond to a viable universe graph.

As expected, I have observed that, for $d$ integer, the minimum energy configuration always corresponds to a $2d$-regular graph with exactly $2d$ edges for every vertex. By the degree sum formula $2e = \sum_{i\ge3} i \ v_i$, with $e$ the number of edges  and $v_i$ the number of vertices of degree $i$, we can derive that these graphs have exactly $dN$ edges. The connectivity details of the ground state graph, like clustering coefficient, number of 4-cycles and others can indeed vary from run to run but they do correspond to finite size effects: the bulk properties convergence to unique values for large values of $N$. 

The three most important quantities that I have measured are the {\it clustering coefficient}, the {\it spectral dimension} and the {\it Hausdorff dimension} of the emergent dynamical graphs.  As already anticipated, the clustering coefficient c \cite{graphrev} is related to the frequency of triangles in the graph: c = 3 x number of triangles / number of connected triples,  which can be also written as c = number of closed triples / number of connected triples. In this last form it can be easily evaluated once the adjaceny matrix $A$ of the ground state graph is known:
\begin{equation}
c = {{\rm Tr} A^3 \over \sum_{ij}A^2_{ij} -Tr A^2} \ .
\label{clust}
\end{equation}

The spectral dimension $d_s$ measures the connectivity of the graph \cite{spectral}, the dimension that a particle moving on the graph would feel.  It is defined via the scaling of the return probability $p_r(t)$ to the initial point after $t$ steps of a random walk on the graph \cite{rwalk}. 
\begin{equation}
p_r (t) \sim t^{-d_s / 2 } \ .
\label{twoj}
\end{equation}
For infinite graphs this scaling relation is valid in the limit $t\to \infty$. For finite $k$-regular graphs, the return time to the initial point is $t=N$ \cite{rwalk} and thus $p_r(t) \to 1$ in the limit $t \to \infty$. The correct scaling is typically found in the intermediate region $1\ll t \le O(N^{4/k})$ where finite size effects are suppressed \cite{rwalk}. To compute the spectral dimension one introduces the degree distribution $d_i$ denoting the number of edges at each vertex $i$. The matrix
\begin{equation}
M_{ij}  = {1\over d_i} A_{ij} \ ,
\label{specdima}
\end{equation} 
represents the transition probabilities from vertex $i$ to vertex $j$ in a random walk on the ground state graph. The average over all possible initial points of the return probability of the random walk after $t$ steps is then given by
\begin{equation}
p_r (t) = {1\over N} \ {\rm Tr} M^t \ .
\label{specdimb}
\end{equation}
The best fit of this to a power $t^{-d_s / 2 }$ over the appropriate $t$-interval defines then the spectral dimension $d_s$.

Finally, the intrinsic Hausdorff dimension $d_H$ measures how the graph volume scales with distances ${\cal D}$ on the graph \cite{graphrev},
\begin{equation}
{\cal D} \sim N^{1/d_H} \ .
\label{twok}
\end{equation}
Of course, this depends on the exact definition of the graph distance. Here I will follow \cite{jonsson}
and take ${\cal D} =\langle D \rangle$ as the average distance among vertices of the graph. This can be also computed from the adjacency matrix $A$ of the ground state graph. Indeed, the minimal graph distance between two vertices $i$ and $j$ is given by the smallest integer power ${\ell}_{ij}$ such that $\left(A^{{\ell}_{ij}}\right)_{ij} \ne 0$. The average distance on the graph is thus given by
\begin{equation}
\langle D \rangle = {1\over N(N-1)} \sum_{ij} {\ell}_{ij} \ .
\label{haus}
\end{equation}
Let me stress that this is an intrinsic property of the graph that does not need an embedding in an extrinsic Euclidean space to be defined and measured. It is the dimension that would be measured with clocks and rods by "inhabitants" of the graph. This dimension is simply extracted as the best fit of (\ref{haus}) to a power of $N$ as in (\ref{twok}). 

For large values of N the convergence to the ground states slows down considerably and so does the computing time needed to evaluate the quantities of interest. This difficulty is compounded by the increasing probability of ending up in local minima of the energy function. These difficulties, combined with the limits of the available computational resources have bounded the accessible size of graphs to vertex numbers of up to a maximum of $N=250$ ($N(N-1)/2= 31125$ possible links). 

The first important observation is the strong evidence for a first quantum phase transition at $d=1$ ($J=1/3$), a quantitative change in behaviour of the ground state at zero temperature as a function of the model coupling constant \cite{sachdev}. When $d=1$ is approached from above the dynamical graphs start to break up into many disconnected components, as shown in Fig. 1 for $N=120$.

\begin{figure}
\includegraphics[width=8cm]{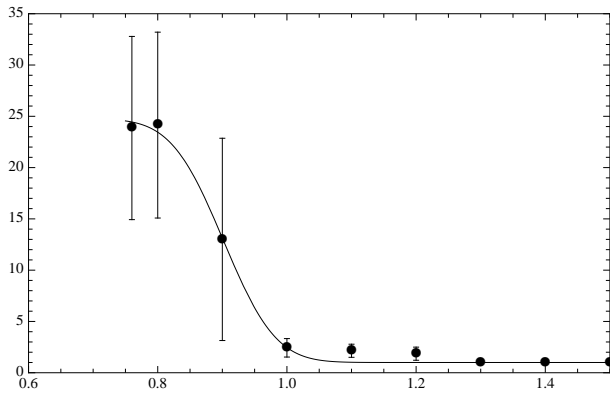}
\caption{\label{fig:Fig1} The average number of disconnected components of dynamical space-time graphs as a function of the coupling constant $d$ for $N=120$. All graphs with values between $d=0.75$ and $d=1.25$ have connectivity 2, corresponding to a one-dimensional Ising model for the space-time vertices. This supports the existence of a quantum phase transition at the lower critical dimension $d=1$ of the Ising model. }
\end{figure}

This happens from $d=1.25$, where some vertices of the graph start to have connectivity 2 (corresponding to one dimension). Of course, the number of connected components diverges like $N$ for $d\le 0.75$, where some vertices begin to have connectivity 1 (i.e. less than one dimension). The interesting region, however, lies between $d=1.25$ and $d=0.75$, where the connectivity is always 2 (corresponding to the one-dimensional Ising model) for all vertices and the $N$-dependence is much less pronounced. Here, there is a clear evidence of a quantum phase transition at $d=1$, where the number of disconnected components increases dramatically even at these small finite sizes. Note that this happens even if the fictitious energy-descent dynamics is chosen to favour aligned space-time spins, as in eq. (\ref{twof}). For a completely symmetric dynamics the proliferating disconnected components have also disordered space-time spins. 

In order to understand the nature of this phase transition it is helpful to consider the model (\ref{twob}) as a standard nearest-neighbours Ising model with dynamical dimensionality $d$. As is well known \cite{zin}, the Ising model has a lower critical dimension $d_{\rm lc} = 1$ below which it is always disordered (at zero temperature). At this lower critical dimension domain walls lower the energy and it becomes thus energetically favourable for the system to create many regions of spins of different signs, eventually completely disordering the system. In the present model with dynamical edges, however, there is an alternative way to lower the energy, namely simply breaking up the links instead of creating domain walls. This is exactly what is happening. This first quantum phase transition at $d=1$ is thus associated with the lower critical dimension of the Ising model. 

Dynamical graphs, emerging as ground states of (\ref{twob}), are low-clustering graphs. Their clustering coefficients scale as $1/N$, as is the case for random graphs and generalized random graphs. Any additional power posited in the $N$-dependence resulted in statistically non-significant parameters of the non-linear regression. In the following table, the clustering coefficients $c_{dg}$ of dynamical graphs and their regression (as a function of N) standard deviations are reported together with the clustering coefficients of generalized random graphs with the same, uniform degree distribution, obtained from (\ref{grapha}). 

\begin{center}
    \begin{tabular}{ | p{1cm} | p{1cm} | p{1cm} | p{1cm} | p{1cm} |}
    \hline
    d & 2 & 3 & 4 & 5 \\ \hline
    $N c_{dg}$ & 2.4 & 4.52 & 6.27 & 8.39 \\ \hline
    $\sigma_{Nc}$ & 0.15 & 0.09 & 0.22 & 0.17 \\ \hline
    $N c_{grg}$ & 2.25 & 4.17 & 6.125 & 8.1 \\ \hline
    \end{tabular}
\end{center}

The two values are very close, although the values for dynamical graphs are slightly higher. This may be the result of the limited number of vertices in the numerical simulations. In any case it is evident that clustering is a finite size effect in dynamical graphs.

In Fig. 2 I plot together the spectral and Hausdorff dimensions as a function of the bare dimension $d$, the coupling constant of the model. As expected, the spectral dimension $d_s$ coincides with the bare dimension $d$. Not so for the Hausdorff dimension $d_H$ though. 

\begin{figure}
\includegraphics[width=8cm]{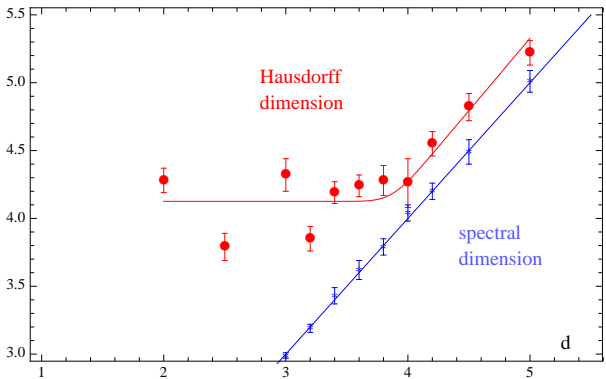}
\caption{\label{fig:Fig. 2} spectral and Hausdorff dimensions as a function of the coupling constant $d$, showing the decoupling of these two dimensions at $d=4$. }
\end{figure}

Indeed, Fig. 2 highlights the evidence for a second quantum phase transition at $d=4$ ($J=1/15$). Although the results are slightly higher than expected, the indication is that, for $d\ge 4$ Hausdorff, spectral and bare dimension coincide while, in the range $1 < d < 4$ the spectral dimension is lower than the Hausdorff dimension, which is fixed at the value $d_H=4$, the required value to describe our observed space-time. The difference $d_H-d_s$ can then be considered as a geometric disorder parameter for this quantum phase transition and is shown in Fig. 3. 

\begin{figure}
\includegraphics[width=8cm]{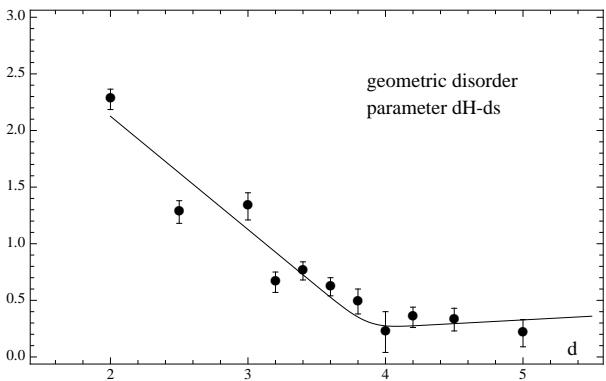}
\caption{\label{fig:Fig. 3} the geometric disorder parameter $d_H$-$d_s$ as a function of the coupling constant $d$. }
\end{figure}

There is actually a plausible explanation of the physical origin of the special role of the dimension 4. To this end let me consider again the interpretation of model (\ref{twob}) as a standard nearest-neighbours Ising model with dynamical dimensionality $d$. The crucial point is that the Ising model has also an upper critical dimension $d_{\rm uc}$ beyond which fluctuations can be neglected altogether and mean field theory becomes exact \cite{zin}. For the Ising model this upper critical dimension turns out to be $d_{\rm uc} =4$. 

It is often stated that hyperscaling fails above the upper critical dimension. This is because, in the standard 
approach to finite size scaling \cite{bre, bin}, one considers the correlation length as bounded by the system size $L$. It has been shown \cite{berche}, however that hyperscaling can be fully restored if this assumption that the finite size correlation length be bounded by the length scale is relaxed. The correct scaling is 
\begin{eqnarray}
\xi &&\propto L^{d/d_{uc}} \ , \qquad d > d_{uc} \ ,
\nonumber \\
\xi &&\propto L \ , \qquad \qquad d \le d_{uc} \ .
\label{hyperone}
\end{eqnarray}
This is because, in finite size samples, the correct correlation length is not commensurate with the physical length above the upper critical dimension \cite{berche}. It coincides with the physical length below the upper critical dimension but differs above it. One can now define also two volumes, $V=L^d$ and the effective volume  $V_{\rm eff}=\xi ^{d_{uc}}$. These, instead, coincide above the upper critical dimension but differ below it. An alternative, but equivalent way to represent the incommensurability found in \cite{berche} when the number of spins, rather than the geometric size, is fixed, is to consider the effective volume as the volume actually dynamically occupied by the system in any number of dimensions. In terms of this effective volume, the physical length scales as:
\begin{eqnarray}
L &&\propto V_{\rm eff}^{1/d} \ , \qquad \qquad d > 4 \ ,
\nonumber \\
L &&\propto V_{\rm eff}^{1/4}\ , \qquad \qquad d \le 4 \ ,
\label{hypertwo}
\end{eqnarray}
where I have already explicitly used $d_{uc}=4$ for the Ising model. This is exactly what I have found for the Hausdorff dimension of model (\ref{twob}). In other words, the quantum phase transition of (\ref{twob}) at $d=4$ (or equivalently at $J=1/15$) is due to the change in critical behaviour of the Ising model at its upper critical dimension. For $d \ge 4$ the model is thus in a mean field phase with $d=d_s=d_H$, for $d <4$, instead fluctuations become important and the model is in a fluctuations dominated phase in which $d_H=4$ and $d_s<d_H$. The dimension 4 of the universe would thus be simply related to the fact that 4 is the upper critical dimension of the Ising model. Note that a spectral dimension lower than the Hausdorff dimension is also exactly what is predicted by the CDT approach \cite{spec}. On the graph sizes accessible to numerical simulations, I have not been able to observe a scale-dependence of the spectral dimension, as predicted by the CDT approach \cite{spec}. It seems, however natural to conjecture that the quantum phase transition at $d=4$ corresponds to an infrared-stable critical point, especially in view of the fact that the Gaussian model is an infrared fixed point of the Ising model at $d=4$. In this case the model (\ref{twob}) would predict the large-scale dimension 4 of the universe and realize at the same time the asymptotic safety scenario of quantum gravity \cite{triang}. The other quantum phase transition at $d=1$, in fact, would necessarily imply the existence of an ultraviolet-stable fixed point either at $d=1$ or in the intermediate region $1<d<4$. In both cases the spectral dimension would be scale-dependent. The latter case would correspond to a universe of lower dimensionality at short distances. The former, even more interesting case, would imply that there is no space-time at all at short distances, only fluctuating information bits, something like the ``quarks of quantum gravity". The fact that $T=0$ is an ultraviolet fixed point of the one-dimensional Ising model \cite{cardy} supports this picture. 

To conclude this section I would like to discuss the role of quantum mechanics in the present model. Indeed, the Hamiltonian (\ref{twob}) appears at first sight as a purely classical one. But so is also the Hamiltonian of the original $2D$ Kazakov random lattice Ising model \cite{kazakov}. Kazakov's seminal insight was that a new quantum degree of freedom, the fluctuation of the metric, i.e. $2D$ quantum gravity, emerges at the critical point in the local limit. Indeed, it is known since the old days of stochastic quantization \cite{pawu, huf} that the Euclidean quantum partition function of a field theory can be obtained as the thermal partition function of the corresponding classical system coupled to a fictitious thermal bath. The main idea of the present model is essentially the same. The difference lies in the fact that, here, the critical points for which I have given first evidence are two and both lie at zero temperature. When taking the $T\to 0$ limit of the partition function corresponding to (\ref{twob}), the configuration sum reduces to a sum over the many vertex and graph configurations of the same minimal energy. If no transverse magnetic fields are applied, model (\ref{twob}) is identical at the classical and quantum levels since all $\sigma ^3$ Pauli matrices attached to vertices and edges commute. The graph configurations of minimal energy that survive in the $T=0$ partition function can thus be taken to represent the residual zero-temperature quantum fluctuations. The idea is that $4D$ quantum gravity emerges when these become critical in the IR at $d=4$. The other, UV critical point at $d=1$ would then represent the quantum behaviour of the theory at very small scales, where space-time dissolves into random information bits. Note that the Hamiltonian (\ref{twob}) represents one of the typical paradigms for quantum phase transitions \cite{sachdev}. Typically, a Hamiltonian $H=H_0 + J H_1$ can display non-analiticity in its ground state if $H_0$ and $H_1$ commute, exactly as in the present model, leading to level-crossings at particular values of $J$ while the wave functions remain the same \cite{sachdev}.

\section{Topological expansion}
In this section I will focus on the finite-temperature behaviour of the model (\ref{twob}). As stressed in the introduction, here temperature cannot be thought of in terms of time averages of microstates, since there is no time yet. Rather, it has to be considered as a further external parameter that assigns non-vanishing probabilities to graphs different than the ground state according to their energies. 

I will adopt again a technique that is used in the analysis of neural networks \cite{neural}. This consists in promoting the deterministic network dynamics (\ref{twof}) and (\ref{twog}) to a stochastic rule for the probabilities of assuming the binary values, 
\begin{eqnarray}
{\rm Prob} \left( s_i (t+1) = +1 \right) = f \left( h_i (t) \right) \ ,
\nonumber \\
{\rm Prob} \left( w_{ij} (t+1) = +1 \right) = f \left( h_{ij} (t) \right) \ ,
\label{fivea} 
\end{eqnarray}
where
\begin{equation}
f(h) = {1 \over 1+ e^{-2 \beta h}} \ ,
\label{fiveb}
\end{equation}
is the Fermi function at temperature $T=1/\beta$. I will now show that the equilibrium reached by this stochastic update rule is the Boltzman distribution corresponding to the energy function (\ref{twob}). Therefore, assuming an equilibrium configuration, the two above rules will provide coupled mean field equations for mean space-time "magnetization" $<s>$ and the average degree $<k>$ of the graph vertices. This defines a temperature-dependent effective dimension of the universe graph. I don't expect mean field theory to be accurate enough to derive if the two quantum phase transitions are isolated or, rather, the endpoints of continuous phase transitions as a function of the temperature. However, it should be accurate enough for high connectivities $<k>$. 

Let me denote by $F_s \left( s_i \to -s_i \right) $ the probability of a vertex spin flip and by $F_w \left( w_{ij} \to 1-w_{ij} \right)$ the probability of a link spin flip. As a consequence of (\ref{fivea}) these are given by
\begin{eqnarray}
F_s \left( s_i \to -s_i \right) &&= {e^{-\beta h_i s_i} \over 2 \ {\rm cosh} \left( \beta h_i s_i \right)} \ ,
\nonumber \\
F_w \left( w_{ij} \to 1-w_{ij} \right) &&= {e^{-\beta h_{ij} s_{ij} }\over 2 \ {\rm cosh} \left( \beta h_{ij} s_{ij} \right)} \ ,
\label{fivec} 
\end{eqnarray}
where I have introduced $s_{ij} = 2 (w_{ij}-1/2)$. Let me denote by $D\left( \dots s_i \dots w_{ij} \dots \right) $ the probability distribution for activity patterns of all vertex and link spins. Due to the vanishing of the diagonal terms, $w_{ii}=0$, and symmetry $w_{ij}=w_{ji}$ of the link variables I will consider $D$ as function of $w_{ij}$ for $i<j$ only. 
In equilibrium (denoted by "e"), the distribution $D_e$ must satisfy the conditions 
\begin{eqnarray}
&&F_s \left( s_i \to -s_i \right) D_e\left( \dots s_i \dots w_{ij} \dots \right) \nonumber \\
&&= F_s \left( -s_i \to s_i \right) D_e\left( \dots -s_i \dots w_{ij} \dots \right) \ ,
\nonumber \\
&&F_w \left( w_{ij}=0 \to w_{ij}=1\right) D_e\left( \dots s_i \dots w_{ij}=0 \dots \right) \nonumber \\
&&= F_w \left( w_{ij}=1 \to w_{ij}=0\right) D_e\left( \dots s_i \dots w_{ij}=1 \dots \right) \ .
\label{fived} 
\end{eqnarray}
Using (\ref{fivec}) one obtains readily
\begin{eqnarray}
{D_e\left( \dots s_i \dots w_{ij} \dots \right) \over D_e\left( \dots -s_i \dots w_{ij} \dots \right)} = e^{2\beta h_i s_i} \ ,
\nonumber \\
{D_e\left( \dots s_i \dots w_{ij}=1 \dots \right) \over D_e\left( \dots -s_i \dots w_{ij}=0 \dots \right) }= e^{2\beta h_{ij}} \ .
\label{fivee}
\end{eqnarray}
Finally, one easily verifies that these conditions are met by the equilibrium probability distribution
\begin{equation}
D_e\left( \dots s_i \dots w_{ij} \dots \right) = {1\over Z}\ e^{-\beta H\left( \dots s_i \dots w_{ij} \dots \right)} \ ,
\label{fivef}
\end{equation}
where $H$ is the energy function ({\ref{twob}) and $Z$ the corresponding partition function. The factor of 2 in the exponent of (\ref{fivee}) arises since all fundamental variables, both vertex spins $s_i$ and link spins $w_{i<j}$ appear twice in the energy function, the latter due to the symmetry $w_{ij}=w_{ji}$. The equilibrium distribution is thus the Boltzmann distribution corresponding to the energy function ({\ref{twob}).

Having established that (\ref{fivea}) are the probabilities in the equilibrium Boltzmann distribution at inverse temperature $\beta$ I can use them to compute the expectation values of both vertex and link spins in thermal equilibrium. Let me begin by the vertex spins. In this case, the expectation value is 
\begin{equation}
\langle s_i \rangle = +1 f\left( h_i \right) -1 f\left( -h_i \right) = {\rm tanh} \left( \beta \sum_{j\ne i} w_{ij} s_j \right) \ .
\label{fiveg}
\end{equation}
In the mean field approximation, which becomes exact in large number of dimensions, I can bring averages $\langle\dots \rangle$ inside non-linear functions. With the usual homogeneity Ansatz $\langle s_i \rangle = \langle s \rangle $, independent of $i$, I obtain
\begin{equation}
\langle s \rangle = {\rm tanh} \left( \beta \langle k \rangle \langle s\rangle \right) \ ,
\label{fiveh}
\end{equation}
where I have introduced the mean degree $\langle k_i \rangle $ at vertex $i$, 
\begin{equation}
\langle k _i \rangle =\sum_{j\ne i} \langle w_{ij} \rangle \ ,
\label{fivei}
\end{equation} 
and assumed that it is independent of $i$, $\langle k_i \rangle = \langle k \rangle $, $\forall i$. 

One proceeds similarly for link spins. In the mean field approximation, the thermal average is given by
\begin{equation}
\langle w_{ij} \rangle = f\left( h_{ij} \right) = f \left( {1\over 2} \langle s \rangle^2 - J \langle k \rangle + J \langle w_{ij} \rangle \right) \ .
\label{fivej}
\end{equation}
In this case, however, I am not interested in homogeneous solutions in the link variables, but, rather, only the average degree (\ref{fivei}) is expected to be independent of the vertex label. I will therefore sum the left hand side of (\ref{fivej}) over all $j \ne i$ to obtain 
\begin{equation}
\langle k \rangle = \sum_{j\ne i} f \left( {1\over 2} \langle s \rangle^2 - J \langle k \rangle + J \langle w_{ij} \rangle \right) \ .
\label{fivek}
\end{equation}
At high temperatures one can expect the thermal average of the link spins to approach its limiting value 1/2 from below. I will thus approximate inside the Fermi function $\langle w_{ij} \rangle \simeq 1/2 - \epsilon_{ij}$, $\epsilon_{ij} >0$ and neglect the deviations $\epsilon_{ij}$, keeping however in mind that they are positive. This gives
\begin{equation}
{\langle k \rangle \over N-1} =  f^- \left( {1\over 2}  \langle s \rangle^2 - J \left( \langle k \rangle -{1\over 2}\right) \right) \ ,
\label{fivel}
\end{equation}
where the superscript "-" means that 0 must be approached from below in the Fermi function. Equations (\ref{fiveh}) and (\ref{fivel}) constitute the coupled mean field equations for "space-time magnetization" and mean degree, respectively. They describe topological fluctuations of discrete space-time graphs and their connectivity at inverse temperature $\beta$ and should provide a correct picture of these emerging universe graphs as long as $\langle k \rangle \gg 1$. 

To solve these equations I will introduce the new variable $x = (1/2) \langle s \rangle^2 - J \left( \langle k \rangle -(1/2)\right)$ and the new linear function 
\begin{equation}
g(x) = {1\over N-1} \left( {1\over 2J} \langle s \rangle ^2 +{1\over 2} \right) - {1\over J(N-1)} x \ .
\label{fivem}
\end{equation}
and rewrite the mean field equations as
\begin{eqnarray}
\langle s \rangle &&= {\rm tanh} \left( \beta \langle k \rangle \langle s\rangle \right) \ ,
\label{fiven} \\
g(x) &&= f^- (x) \ .
\label{fiveo}
\end{eqnarray}
When $\beta \langle k \rangle \le 1$, eq. (\ref{fiven}) has the unique solution $\langle s \rangle = 0$. This is the case in the high temperature regime $\beta \to 0$ (for finite $\langle k \rangle$). In this limit, the Fermi function in (\ref{fiveo}) reduces to its limiting value 1/2, giving the solution $g(x) = 1/2$ or $\langle k \rangle= \langle k \rangle_{\rm max} = (N-1)/2$, which is indeed finite. This is the information phase of the model, a hot soup of information bits with no space-time interpretation. When  $\beta \langle k \rangle > 1$, instead, two solutions of 
(\ref{fiven}) exist, the same solution $\langle s \rangle = 0$ as before and a solution with finite $\langle s \rangle >0$. Only the latter, however, is stable, as is well known from the mean field theory of the Ising model \cite{zin}. In the low temperature regime $\beta \to \infty$ (with $\langle k\rangle \langle s\rangle >0$) the stable solution becomes $\langle s \rangle = 1$. In this limit, the Fermi function approaches a Heaviside function at $x=0$, always crossing the linear function $g(x)$ when $x\to 0^-$, which amounts to $\langle k \rangle \to \langle k \rangle_{\rm min} = 2d $. This is the topology phase of the model, in which the ground state of the model is a low-clustering universe graph with effective Hausdorff dimension $d_{\rm eff} = \langle k \rangle /2 \ge 4$. Between the two phases there is a ferromagnetic phase transition in which space-time and the discrete universe emerge spontaneously. The exact position of the phase transition can be determined by solving numerically the system of equations (\ref{fiven}) and (\ref{fiveo}). For example, for $d=4$ it is located at $T_{\rm cr}= \langle k \rangle _{\rm cr} = 0.468 N$. 

In the topology phase of the model the average connectivity $\langle k \rangle $ of the universe increases with decreasing temperature. This corresponds to a topological expansion in which the universe actually ``unfolds" rather than expands. There is no big bang, space-time emerges collectively in a phase transition as a very tight ball of ``hyperconnected" points with scaling factor $D_{\rm min} \propto N^{2/\langle k\rangle_{\rm cr}}$. Because of the very high connectivity at the phase transition,  $\langle k \rangle_{\rm cr} =O(N)$ this is a finite quantity in the limit $N\to \infty$. Note also that there is no singularity at the phase transition, the universe graph emerges with an average distance of O(1). 

For very high and very low temperatures it is possible to derive how the average distance on the universe graph scales with the temperature. For very high temperatures, $\beta \to 0$ one can expand the exponential in the Fermi function, for very low temperatures $\beta \to \infty$, instead, one can replace the Fermi function (\ref{fiveo}) with ${\rm exp}(x)$ (bear in mind that $x$ is a negative quantity). This gives
\begin{eqnarray}
\langle D\rangle &&\propto D_{\rm min} \ N^{2J\over T} \ , \ \ \qquad  T\to \infty \ ,
\label{fivep} \\
\langle D\rangle && \propto D_{\rm max} \ N^{-aT} \  , \qquad T\to 0 \ ,
\label{fiveq}
\end{eqnarray}
with $a=\left( 4J^2/(J+1)^2 \right) {\rm log}\left( 2J(N-1)/J+1\right)$ and $D_{\rm max} \propto N^{1/d}$ for $d>4$ and $D_{\rm max} \propto N^{1/4}$ for $d \le 4$. The important point is that, during topological expansion, the average distances in the universe scale exponentially with the temperature (or its inverse). A tiny decrease in temperature causes a large increase in the ``topological scale factor" of the universe graph. This starts out as an almost completely connected graph and expands exponentially (with temperature) to a regular graph of dimension 4 (if $d\le 4$). 

Let me finally mention a very important point. Mean field theory addresses only the average degree $<k>$ of the universe graph. In the phase of interest, however, $<k> \ll N$ and thus the exact solution will assign non-vanishing probabilities to only approximately $<k>$ of the possible $N(N-1)/2$ edges at every vertex. What is more, when $<k>\ne {\rm integer}$ these non-vanishing probabilities will typically differ from each other, a situation most pronounced for $<k> = 2d + \epsilon$, where I expect some edges with probabilities much smaller than others. Finite temperature is thus a mechanism leading to possible anisotropies in the universe graph.

\section{Topological black holes}
Until now I have concentrated on emerging universe graphs with homogeneous space-time ``magnetization" and vertex degrees, both at $T=0$ and at finite temperature in the mean field limit $\langle k \rangle \gg 1$. In this section, instead, I will explore possibly inhomogeneous configurations that can emerge at generic, lower values of $\langle k \rangle $.

To do so I will consider a Peierls droplet of $n \ll N$ wrong space-time $s_i=-1$ spins in an otherwise homogenous universe of $N$ vertices. In the usual Ising model, the vertex connectivity is fixed and, when it is high enough, these droplets disappear below a finite critical temperature since the entropy of the droplet becomes lower than its internal energy, so that the free energy is dominated by the latter \cite{bon}. This is given by the boundary contribution $E= E_h + \delta$ where $E_h$ is the energy of the homogeneous universe graph and $\delta $ is the number of links between spins of opposite signs. The internal energy is thus minimized by turning all wrong spins so that $\delta \to 0$. This is the standard Peierls argument to explain why all the spins align when lowering the temperature and it is exactly what happens in the homogenous mean field solution discussed in the previous section. 

In the present model, instead, the links are themselves dynamical variables and there is another way to lower the internal energy of the droplet. Indeed, turning one single spin does not lower the internal energy but it actually increases it by $\Delta E= \langle k \rangle -2$, where $\langle k \rangle$ is the mean connectivity at the considered temperature, since for one "corrected" link there are now, on average $\langle k \rangle-1$ new wrong ones. For high enough temperatures, $\langle k \rangle > 2$ and the energy cost is a positive quantity. Of course, if the links are given external parameters, the only way to lower the energy is a simultaneous turnaround of all the wrong spins. Here, however the system could simply severe the links between spins of opposite signs, thereby lowering the internal energy by $\Delta E= -\delta/2 \left( 1+ J(\langle k\rangle-1)\right) $. The resulting configuration is a universe represented by a graph containing a hole within which there is simply no space-time. This hole is represented by a disconnected subgraph for which all vertex spins have value $s_i=-1$. The original universe graph contains a ``Schwarzschild-like" boundary on which the spectral dimension is lower by one than in the rest of the universe graph: this boundary is constituted by all $s_i=+1$ vertices for which a link to a $s_i=-1$ vertex inside the hole has been severed. I will call such a configuration a topological black hole. It is characterized by the absence of space-time inside the hole, by a lowered spectral dimension on its boundary and, at least for integer $d$, by an entropy proportional to the ``area", i.e. the number of vertices on the boundary. For $d=2$ this is the original Peierls argument \cite{zin}. The Peierls argument, however, can be generalized to any integer $d\ge 2$ \cite{bon}, the entropy always scales as the number of vertices on the higher-dimensional contour around the hole, i.e. its ``area".

The question arises if a topological black hole is, on average, a stable configuration for temperatures down to $T=0$. The only way to detect the presence of the topological black hole from within the main universe graph is through the lower spectral dimension on its boundary. And indeed, the lowest-energy instabilities come from vertices on the boundary forming links between themselves. Forming such a link would lower the system energy by $1/2$ via the ferromagnetic vertex term in the energy, on the other side, it would cost energy due to the antiferromagnetic link term. In total, one such link would cost an energy 
\begin{equation}
\Delta E = J(\langle k \rangle-1) -{1\over 2} \ .
\label{sixa}
\end{equation}
Note that forming links to other vertices not on the boundary costs more energy since they possess already more links. Topological black holes are thus, on average, stable configurations against radius connections if (\ref{sixa}) is a positive quantity. Using $J=1/(4d-1)$ this stability condition can be formulated as
\begin{equation}
\langle k\rangle > 2d+{1\over 2} = k(T=0) + {1\over 2} \ .
\label{sixb}
\end{equation}
This shows that topological black holes are typically stable down to very low temperatures. 

\section{Future directions} 
Two further steps are needed to make definitive contact with gravity. First and foremost the renormalization flow of the dimension and the local limit at the IR fixed point have to be derived and it must be shown how the Einstein-Hilbert action arises there. Secondly it remains to show how the local limit admits a causal structure and why the emerging continuous space is Lorentzian. These are the subjects of further investigation.

\begin{acknowledgments}
I would like to thank Luis Alvarez-Gaum\'e and Fernando Quevedo for a critical reading of the manuscript and very helpful comments and Jan Ambj\o{}rn for a very helpful discussion. Many thanks also to R. Kenna and B. Berche for explaining the fine points of their work on hyperscaling in the Ising model. 
\end{acknowledgments}

\end{document}